# Fresnel and Fraunhofer diffraction of a laser Gaussian beam by fork-shaped gratings

Ljiljana Janicijevic and Suzana Topuzoski [*]

Institute of physics, Faculty of natural sciences and mathematics, University "Sts Cyril and Methodius", 1000 Skopje, Republic of Macedonia

*Corresponding author: suzana_topuzoski@yahoo.com

Expressions describing the vortex beams, which are generated in a process of Fresnel diffraction of a Gaussian beam, incident out of waist on a fork-shaped gratings of arbitrary integer charge $p$, and vortex spots in the case of Fraunhofer diffraction by these gratings are deduced. The common general transmission function of the gratings is defined and specialized for the cases of amplitude holograms, binary amplitude gratings, and their phase versions. Optical vortex beams, or carriers of phase singularity with charges $mp$ and $-mp$, are the higher negative and positive diffraction order beams. The radial part of their wave amplitudes is described by the product of $mp$-th order Gauss-doughnut function and a Kummer function, or by the first order Gauss-doughnut function and a difference of two modified Bessel functions, whose orders don't match the singularity charge value. The wave amplitude and the intensity distributions are discussed for the near and far field, in the focal plane of a convergent lens, as well as the specialization of the results when the grating charge $p=0$, i.e. the grating turns from forked into rectilinear. The analytical expressions for the vortex radii are also discussed.

## 1. INTRODUCTION

Optical vortices have attracted an increased interest and have become very important in optical trapping and manipulation of small particles. Among the optical elements which generate optical vortex beams are phase spiral plate [1,2,3], helical axicon [4,5,6], forked grating [7,8,9,10], spiral zone plates [11]. It was recognized by Allen et al. [12] for the first time that a beam with helical wavefront characterized by azimuthal phase $\exp(il\varphi)$, where $\varphi$ is the azimuthal coordinate, $l$ is an integer, posses an orbital angular momentum $l\hbar$ per photon, in its propagation direction.

In [13] the authors analyze hypergeometric laser beams that are generated by complex amplitude composed of four cofactors: the Gaussian exponent, a logarithmic axicon, a spiral phase plate and an amplitude power function with a possible singularity at the origin of the coordinates.

Forked interferograms and gratings belong to the class of diffractive optical elements which generate optical vortices. Based on their experimental experience, the authors in [7] gave the first report on the production of optical fields with wavefront dislocations of various orders by the use of holograms possessing fork-shaped dislocations.

An effort for theoretical description of the Gaussian wave field transformation by the forked hologram has been done in [8]. The authors define the transmission function of the forked hologram, and multiplying it by the Gaussian incident beam, calculate the field immediately after the hologram. It consists of three components-one chargeless and two with induced phase singularities of opposite charges. But, without using the method of solving the diffracted wave integrals, any information about the amplitude form and the spatial configuration of the field components isn't obtained. The proposed description of the charged components as Laguerre-Gaussian beams is based only on the experimental far field registration of the ring-like transverse amplitude (intensity) profiles and on their singularity detection by interferometric methods.

The situation is similar with the results in [9], where the Fraunhofer diffraction of a plane wave by forked gratings is treated. The authors solved the problem in cylindrical coordinates, performing the integration over the azimuthal variable only. By doing so, it is shown that there exists a generation of beams with phase singularities, but the amplitude (intensity) profile of such beams remains unknown. It makes impossible the calculation of the vortex size through the radial intensity distribution, so important in experimental applications.

Computer generated holograms have been used to create hollow beams or optical vortices, which were used for guiding of cold atoms [14] trapped in the dark region surrounded by a repulsive dipole potential wall (named 'blue-detuned" traps). Knowing the radial intensity distribution $I(r)$ of the diffracted field is very important, because the optical dipole potential spatial variation $U(r)$ depends upon the beam intensity profile [15,16] as

$$U(r) \propto \ln[1 + I(r)/F],$$

where the value of $F$ depends on the saturation intensity, the natural width of the atomic transition and the frequency detuning from the atomic resonance.

Therefore, the forked gratings and holograms, as optical elements which generate optical vortices, need a complete solution for the diffracted wave and intensity distribution.

In [17] the authors have analytically derived the expressions for diffraction of nondiverging ("nondiffracting") Bessel beams by fork-shaped grating and rectilinear grating. The diffracted wave field consists of a nondiverging, amplitude reduced beam of the same topological charge and order as the incident beam (as a zero diffraction order), and higher diffraction order vortex beams whose charge is a sum of the incident beam charge and that of the forked grating multiplied by the diffraction order. The amplitude of the higher diffraction order beams is described by a sum of Gauss hypergeometric functions.

In this article we solve the problem of an out of waist incident Gaussian beam on a grating with a fork-shaped singularity, identified by the number of its "internal teeth" (Fig. 1 a), including the case of its absence (Fig. 1 b). The wave and the intensity distribution are expressed through the Bessel functions whose order don't match the singularity order of the beam, they are discussed in the near and far field, as well as the specialization of the results for the case of rectilinear grating. Also, the analytical expressions for the Fraunhofer diffraction of a Gaussian beam, incident on a forked grating with its waist, in the focal plane of a spherical convergent lens are derived and discussed. The vortex radii and the behavior of the beam near and far from the vortex centers are described by analytical formula.

## 2. THE TRANSMISSION FUNCTION OF THE FORK-SHAPED GRATING (FSG)

The transmission function of the fork-shaped hologram can be realized by the photographic registration and reduction of the irradiance distribution of the interference between a slightly inclined plane wave with amplitude $A_1$ and a wave with singularity of order $p$ in azimuthal direction of its phase wave front, with amplitude $A_2$. The binary versions of the grating are computer generated and photoreduced transparences.

The general transmission function of the fork-shaped grating in cylindrical coordinate system is defined as

$$T(r,\varphi) = \sum_{m=-\infty}^{\infty} t_m \exp\left[-im\left(\frac{2\pi}{D}r\cos\varphi - p\varphi\right)\right]$$
$$= t_0 + \sum_{m=1}^{\infty} t_m \exp\left[-im\left(\frac{2\pi}{D}r\cos\varphi - p\varphi\right)\right] + \sum_{m=1}^{\infty} t_{-m} \exp\left[im\left(\frac{2\pi}{D}r\cos\varphi - p\varphi\right)\right] \quad (1)$$

with $(r,\varphi)$ being the coordinates in grating's plane, $p$ is an integer showing the number of the internal "teeth" of the forked grating. When $p=0$ the transmission function turns into rectilinear grating. The constant $D$ is a period of the rectilinear grating, and it plays the same role for the FSG far from its pole.

The specifications of the transmission coefficients $t_{\pm m}$ depend on the type of the grating.

For amplitude grating with sinusoidal transmission, equation (1) contains only terms with $m=-1,0,1$: $t_1 = t_{-1} = t/2$. $t_0$ and $t$ are transmission coefficients of the film transparency, connected to $A_1$ and $A_2$ in the following way: $t_0 = A_1^2 + A_2^2$; $t = 2A_1 A_2$.

The transmission coefficients of binary amplitude gratings are: $t_0 = 1/2$, $t_{\pm m = \pm(2m'-1)} = \mp i/\pi(2m'-1)$, $t_{\pm m = \pm 2m'} = 0$ ($m'=1,2,3,...$).

For the phase holograms, the transmission coefficients are defined as:

$t_0 = J_0(k\beta)\exp(ik\alpha)$, $t_{\pm m} = (-i)^{\pm m} J_{\pm m}(k\beta)\exp(ik\alpha)$, ($m=1,2,..$), with $k\alpha$ being the bias phase retardation, $k\beta$ being the amplitude of the phase modulation and $J_{\pm m}(k\beta)$ denotes Bessel function of integer order.

For binary phase grating with rectangular profile, the coefficients are the following [18]:

$t_0 = \chi \exp(ik\alpha)\cos(k\beta)$; $t_{\pm m = \pm(2m'-1)} = \pm \chi \exp(ik\alpha)\frac{2}{\pi}\frac{1}{(2m'-1)}\sin(k\beta)$; $t_{\pm 2m'} = 0$; ($m'=1,2,3,...$),

where $\chi$ defines the coefficient of transmission of the phase layer (which is equal to one for ideal transmission), while $k\alpha$ and $k\beta$ have the same meaning as for the phase hologram. For the case $\chi=1$ and $k\beta=\dfrac{\pi}{2}$ it follows that: $t_0=0$, $t_{\pm(2m'-1)}=\pm\dfrac{2}{\pi}\dfrac{1}{(2m'-1)}\exp(ik\alpha)$, $t_{\pm 2m'}=0$.

The pole of the coordinate system is located on the bottom of the internal grating teeth, as it is shown in Fig. 2.

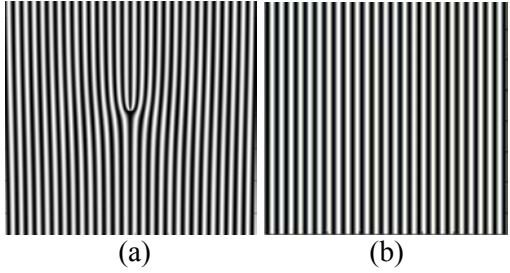

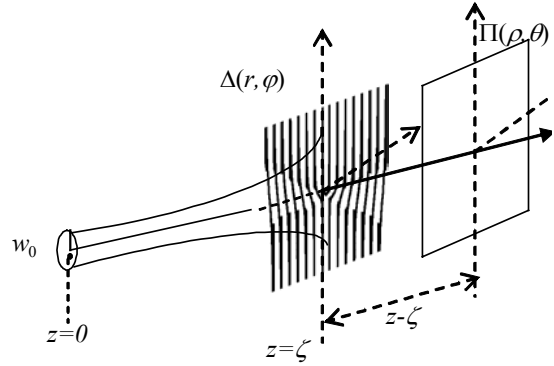

Fig. 1. (a) Forked grating with $p=2$ and (b) rectilinear grating when $p=0$.

Fig. 2. The geometry of the problem.

## 3. THE FRESNEL DIFFRACTION OF A GAUSSIAN LASER BEAM BY THE FSG

The study of the Fresnel diffraction gives an opportunity to discuss the spatial form and divergence qualities (or diffractive spreading) of the beams generated by the FSG.

To avoid the choice of infinite limits of the diffracting element area of the integration, it is convenient the incident wave field to be in the form of a monochromatic Gaussian beam of wavelength $\lambda$. At a distance $\zeta$ from its waist, the incident beam is represented as

$$U^{(i)}(r,\varphi,\zeta)=\frac{q(0)}{q(\zeta)}\exp\left[-ik\left(\zeta+\frac{r^2}{2q(\zeta)}\right)\right], \qquad (2)$$

where $k=2\pi/\lambda$ is propagation constant, $q(\zeta)=\zeta+ikw_0^2/2$ is the beam complex parameter, $w_0$ is the beam waist radius. To the beam's complex curvature

$$\frac{1}{q(\zeta)}=\frac{1}{R(\zeta)}-\frac{2i}{kw^2(\zeta)}, \qquad (3)$$

a real on-axial radius of curvature $R(\zeta) = \zeta\left[1 + \left(kw_0^2/2\zeta\right)^2\right]$ is assigned, while $w(\zeta) = w_0\left[1 + \left(2\zeta/kw_0^2\right)^2\right]^{1/2}$ is the beam transverse amplitude profile radius.

The fork-shaped grating is situated a distance $z=\zeta$ from the beam waist (Fig. 2). The incident beam axis is passing through the pole of the grating, situated in the plane $\Delta(r,\varphi)$, and in the same time it plays the role of $z$-axis of the cylindrical system we are working with.

In the observation screen $\Pi(\rho,\theta)$, situated at a distance $(z-\zeta)$ from the grating plane, the wave field in the point $(\rho,\theta,z)$ can be found using the Fresnel-Kirchoff integral [19]

$$U(\rho,\theta,z) = \frac{ik}{2\pi(z-\zeta)} \exp\left\{-ik\left[(z-\zeta) + \frac{\rho^2}{2(z-\zeta)}\right]\right\}$$
$$\times \iint_\Delta T(r,\varphi) U^{(i)}(r,\varphi,\zeta) \exp\left[-i\frac{k}{2}\left(\frac{r^2}{z-\zeta} - \frac{2r\rho\cos(\varphi-\theta)}{z-\zeta}\right)\right] r\, dr\, d\varphi, \qquad (4)$$

where $\Delta$ is the area of the grating.

The insertion of Eq. (1) and Eq. (2) in the diffraction integral (4) yields

$$U(\rho,\theta,z) = \frac{ik}{2\pi(z-\zeta)} \frac{q(0)}{q(\zeta)} \exp\left[-ik\left(z + \frac{\rho^2}{2(z-\zeta)}\right)\right]$$
$$\times \left\{ t_0 \int_0^\infty \int_0^{2\pi} \exp\left[-i\frac{k}{2}\frac{q(z)}{(z-\zeta)q(\zeta)}r^2\right] \exp\left[\frac{ikr\rho}{(z-\zeta)}\cos(\varphi-\theta)\right] r\, dr\, d\varphi \right.$$
$$+ \sum_{m=1}^\infty t_m \int_0^\infty \int_0^{2\pi} \exp\left[-i\frac{k}{2}\frac{q(z)}{(z-\zeta)q(\zeta)}r^2\right] \exp\left[\frac{-ikr\rho}{(z-\zeta)}\sin\theta\sin\varphi\right]$$
$$\times \exp\left[\frac{ikr\cos\varphi}{z-\zeta}\left(\rho\cos\theta - \frac{m\lambda(z-\zeta)}{D}\right)\right] \exp(imp\varphi) r\, dr\, d\varphi$$
$$+ \sum_{m=1}^\infty t_{-m} \int_0^\infty \int_0^{2\pi} \exp\left[-i\frac{k}{2}\frac{q(z)}{(z-\zeta)q(\zeta)}r^2\right] \exp\left[\frac{-ikr\rho}{(z-\zeta)}\sin\theta\sin\varphi\right]$$
$$\left. \times \exp\left[\frac{ikr\cos\varphi}{z-\zeta}\left(\rho\cos\theta + \frac{m\lambda(z-\zeta)}{D}\right)\right] \exp(-imp\varphi) r\, dr\, d\varphi \right\}. \qquad (5)$$

Further, we use the following variable transformations that concern the observation plane

$$\rho\cos\theta \mp \frac{m\lambda(z-\zeta)}{D} = \rho_{\pm m}\cos\theta_{\pm m},$$
$$\rho\sin\theta = \rho_{\pm m}\sin\theta_{\pm m}, \qquad (6)$$

with $\rho_0 = \rho$ and $\theta_0 = \theta$, but

$$\rho_{\pm m}=\sqrt{\rho^2+\left[\frac{m\lambda(z-\zeta)}{D}\right]^2 \mp \frac{2m\lambda\rho(z-\zeta)}{D}\cos\theta};\quad \mathrm{tg}\,\theta_{\pm m}=\frac{\rho\sin\theta}{\rho\cos\theta\mp m\lambda(z-\zeta)/D}. \qquad (7)$$

The pairs of the newly introduced variables $(\rho_m,\theta_m)$ and $(\rho_{-m},\theta_{-m})$ in the observation plane $\Pi(\rho,\theta)$ play the role of plane polar coordinates, related to the poles $C_m(m\lambda(z-\zeta)/D,0)$ and $C_{-m}(m\lambda(z-\zeta)/D,\pi)$ respectively (Fig. 3).

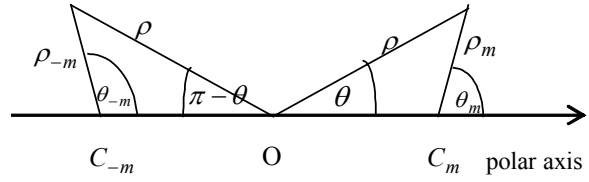

Fig. 3. Geometrical illustration of the transformations (6).

The solutions of the integrals over the azimuthal variable in Eq. (5)

$$\Phi_0=\int_0^{2\pi}\exp\!\left\{i\frac{k\rho}{(z-\zeta)}r\cos(\varphi-\theta)\right\}d\varphi;\quad \Phi_{\pm mp}=\int_0^{2\pi}\exp\!\left\{i\left[\frac{k\rho_{\pm m}}{(z-\zeta)}r\cos(\varphi-\theta_{\pm m})\pm mp\varphi\right]\right\}d\varphi \qquad (8)$$

are easily found by the use of the Jacoby-Anger identity, known from the theory of Bessel functions [20], and are represented as

$$\Phi_0=2\pi J_0\!\left(\frac{k\rho}{z-\zeta}r\right);\quad \Phi_{\pm mp}=2\pi J_{mp}\!\left(\frac{k\rho_{\pm m}}{z-\zeta}r\right)\exp[imp(\pi/2\pm\theta_{\pm m})]. \qquad (9)$$

The insertion of this result in the expression (5) gives

$$U_m(\rho_m,\theta_m,z)=\frac{ik}{(z-\zeta)}\frac{q(0)}{q(\zeta)}\exp\!\left[-ik\!\left(z+\frac{\rho^2}{2(z-\zeta)}\right)\right]\Bigg\{t_0\int_0^{\infty}J_0\!\left(\frac{k\rho r}{z-\zeta}\right)\exp\!\left[-\frac{ik}{2}\frac{q(z)}{(z-\zeta)q(\zeta)}r^2\right]rdr$$

$$+\sum_{m=0}^{\infty}t_m\exp\!\left[imp\!\left(\frac{\pi}{2}+\theta_m\right)\right]\int_0^{\infty}\exp\!\left[-\frac{ik}{2}\frac{q(z)}{(z-\zeta)q(\zeta)}r^2\right]J_{mp}\!\left(\frac{k\rho_m r}{z-\zeta}\right)rdr$$

$$+\sum_{m=0}^{\infty}t_{-m}\exp\!\left[imp\!\left(\frac{\pi}{2}-\theta_{-m}\right)\right]\int_0^{\infty}\exp\!\left[-\frac{ik}{2}\frac{q(z)}{(z-\zeta)q(\zeta)}r^2\right]J_{mp}\!\left(\frac{k\rho_{-m}r}{z-\zeta}\right)rdr\Bigg\}. \qquad (10)$$

Therefore, the total diffracted field is represented as a sum of zero-th diffraction order and higher diffraction orders (positive and negative)

$$U(\rho,\theta,z)=U_0(\rho,\theta,z)+\sum_{m=1}^{\infty}U_m(\rho_m,\theta_m,z)+\sum_{m=1}^{\infty}U_{-m}(\rho_{-m},\theta_{-m},z), \qquad (11)$$

with the zero diffraction order wave amplitude

$$U_0(\rho,\theta,z) = \frac{ik}{(z-\zeta)} \frac{q(0)}{q(\zeta)} t_0 \exp\left[-ik\left(z + \frac{\rho^2}{2(z-\zeta)}\right)\right] Y_0(\rho), \quad (12)$$

and the higher diffraction orders wave amplitudes

$$U_{\pm m}(\rho_{\pm m},\theta_{\pm m},z) = \frac{ik}{(z-\zeta)} \frac{q(0)}{q(\zeta)} t_{\pm m} \exp\left[-ik\left(z + \frac{\rho^2}{2(z-\zeta)}\right)\right] \exp\left[imp\left(\frac{\pi}{2} \pm \theta_{\pm m}\right)\right] Y_{\pm m}(\rho_{\pm m}). \quad (13)$$

The integrals over the radial variable in Eqs. (12) and (13) are denoted as

$$Y_0(\rho) = \int_0^\infty J_0\left(\frac{k\rho r}{z-\zeta}\right) \exp\left[-\frac{ik}{2}\frac{q(z)}{(z-\zeta)q(\zeta)} r^2\right] r\,dr, \quad (14)$$

$$Y_{\pm m}(\rho_{\pm m}) = \int_0^\infty J_{mp}\left(\frac{k\rho_{\pm m} r}{z-\zeta}\right) \exp\left[-\frac{ik}{2}\frac{q(z)}{(z-\zeta)q(\zeta)} r^2\right] r\,dr, \quad (15)$$

and they are the well known integrals of Bessel functions [21]

$$\int_0^\infty J_l(b_0 r) \exp(-a^2 r^2) r^{l+1} dr = \frac{b_0^l}{(2a^2)^{l+1}} \exp\left(-\frac{b_0^2}{4a^2}\right) \quad (\operatorname{Re}\nu > -1, \operatorname{Re} a^2 > 0),$$

$$\int_0^\infty J_\nu(b_{\pm m} r) \exp(-a^2 r^2) r^{\mu-1} dr = \left(\frac{b_{\pm m}^2}{4a^2}\right)^{\nu/2} \frac{\Gamma((\mu+\nu)/2)}{2a^\mu \Gamma(\nu+1)} M\left(\frac{\mu+\nu}{2}; \nu+1; -\frac{b_{\pm m}^2}{4a^2}\right)$$

$$(\operatorname{Re}(\mu+\nu) > 0, \operatorname{Re} a^2 > 0).$$

Replacing their solutions in Eqs. (12) and (13) gives the zero-th diffraction order beam

$$U_0(\rho,\theta,z) = t_0 \frac{q(0)}{q(z)} \exp\left\{-ik\left[z + \frac{\rho^2}{2q(z)}\right]\right\}, \quad (16)$$

and the higher diffraction orders

$$U_{\pm m}(\rho_{\pm m},\theta_{\pm m},z) = t_{\pm m} \frac{q(0)}{q(z)} \frac{\Gamma(mp/2+1)}{\Gamma(mp+1)} \left[\frac{-ikq(\zeta)}{2(z-\zeta)q(z)}\right]^{mp/2} \exp\left\{-ik\left[z + \frac{\rho^2}{2(z-\zeta)} + \left(\frac{1}{R(z)} - \frac{1}{z-\zeta}\right) \frac{\rho_{\pm m}^2}{2}\right]\right\}$$

$$\times (\pm 1)^{mp} \exp\left[imp\left(\frac{\pi}{2} \pm \theta_{\pm m}\right)\right] \rho_{\pm m}^{mp} \exp\left(\frac{-\rho_{\pm m}^2}{w^2(z)}\right) M\left(\frac{mp}{2}, mp+1; -\frac{ikq(\zeta)}{2(z-\zeta)q(z)} \rho_{\pm m}^2\right), \quad (17)$$

where $M(\alpha,\beta;x)$ are confluent hypergeometric or Kummer functions, and the Kummer transformation for the hypergeometric functions [21]

$$M\left(\frac{mp}{2}+1, mp+1; -x\right) = \exp(-x) M\left(\frac{mp}{2}, mp+1; x\right) \quad (18)$$

has been used.

The zerot-th diffraction order beam is an ordinary Gaussian, chargeless beam.

Whereas, the higher diffraction order beams posses phase singularities (with exception when $p=0$), with opposite charges $\pm mp$, indicating opposite helical wavefront chiralities of the beams diffracted in the (+$m$)-th and (-$m$)-th diffraction order, respectively. The radial part of the wave function is represented by the product of a doughnut-Gaussian function of order $mp$ and a confluent hypergeometric or Kummer function of a complex parameter.

Using the relation between Kumer and Bessel functions (see appendix 1), we can rewrite expression (17) in the form

$$U_{\pm m}(\rho_{\pm m}, \theta_{\pm m}, z) = t_{\pm m} \frac{q(0)}{q(z)} \left[\frac{ik\pi}{4Q(z)}\right]^{1/2} (\pm 1)^{mp} \exp\left\{-ik\left[z + \frac{\rho^2}{2(z-\zeta)} + \frac{\rho_{\pm m}^2}{2Q(z)}\right]\right\} \exp\left[imp\left(\frac{\pi}{2} \pm \theta_{\pm m}\right)\right]$$

$$\times \rho_{\pm m} \left[ I_{\frac{mp-1}{2}}\left(\frac{ik}{2Q(z)}\rho_{\pm m}^2\right) - I_{\frac{mp+1}{2}}\left(\frac{ik}{2Q(z)}\rho_{\pm m}^2\right) \right]. \tag{19}$$

Here, we've expressed the argument of the Bessel functions as

$$y_{\pm m} = -\frac{ik}{4}\frac{q(\zeta)}{(z-\zeta)q(z)}\rho_{\pm m}^2 = \frac{ik}{2}\left\{\frac{1}{2}\left[\frac{1}{R(z)} - \frac{1}{(z-\zeta)}\right] - \frac{i}{kw^2(z)}\right\}\rho_{\pm m}^2 = \frac{ik}{2Q(z)}\rho_{\pm m}^2,$$

introducing a "new" complex curvature

$$\frac{1}{Q(z)} = \frac{1}{R'(z)} - \frac{2i}{kw'^2(z)}, \tag{20}$$

whose characteristics are connected to those of the incident Gaussian beam in the following way

$$\frac{1}{R'(z)} = \frac{1}{2}\left[\frac{1}{R(z)} - \frac{1}{z-\zeta}\right]; \quad w'(z) = w(z)\sqrt{2}. \tag{21}$$

If we use the identity $I_\nu(ix) = i^\nu J_\nu(x)$, the diffracted components (19) can be expressed, also, as

$$U_{\pm m}(\rho_{\pm m}, \theta_{\pm m}, z) = t_{\pm m} \frac{q(0)}{q(z)} \left[\frac{ik\pi}{4Q(z)}\right]^{1/2} (\pm 1)^{mp} i^{(mp-1)/2} \exp\left\{-ik\left[z + \frac{\rho^2}{2(z-\zeta)} + \frac{\rho_{\pm m}^2}{2Q(z)}\right]\right\}$$

$$\times \exp\left[imp\left(\frac{\pi}{2} \pm \theta_{\pm m}\right)\right] \rho_{\pm m} \left[ J_{\frac{mp-1}{2}}\left(\frac{k}{2Q(z)}\rho_{\pm m}^2\right) - i J_{\frac{mp+1}{2}}\left(\frac{k}{2Q(z)}\rho_{\pm m}^2\right) \right]. \tag{22}$$

The existence of the terms

$$\rho_{\pm m} \exp\left[-i\frac{k}{2Q(z)}\rho_{\pm m}^2\right] = \rho_{\pm m} \exp\left[-\frac{\rho_{\pm m}^2}{w'^2(z)}\right] \exp\left[-\frac{ik}{2R'(z)}\rho_{\pm m}^2\right] \qquad (23)$$

ensures first-order doughnut-Gaussian modes of the amplitude profile of the beams. Their propagation axes $\rho_m = 0$ and $\rho_{-m} = 0$, or

$$(\rho = m\lambda(z-\zeta)/D, \theta = 0) \text{ and } (\rho = m\lambda(z-\zeta)/D, \theta = \pi), \qquad (24)$$

are axes of zero amplitude (intensity) values. They deviate from the axes of the zero-order beam (z-axis) on both sides for angles $\delta_{\pm m} = \arctan(m\lambda/D)$.

The orders of the Bessel functions in (19) and (22) don't match the topological charge, which differs from the results in [4,5], where the diffraction of a plane wave by computer generated hologram with transmission function of a helical axicon is treated, and the diffraction components are described by Bessel functions of integer order which equals the topological charge value of the vortex beam components. Another difference is that here the argument of the Bessel functions depends on the quadratic term of the radial coordinate, instead linearly. Eqs. (19) and (22) are similar to the case of diffraction of a plane and Gaussian beam by a spiral phase plate with singularity's integer order $n$, presented in detail in [3], where as a result of diffraction, only one diffraction order beam is obtained, propagating along the optical axis (defined by the incident beam axis). In this work, as a result of diffraction of a Gaussian beam by forked grating, a fan of separated diffraction order beams is observed. In the $(\pm m)$-th diffraction order ($m = 1,2,3..$), the beam has topological charge $\pm mp$ and is described by Bessel functions of order $(mp-1)/2$ and $(mp+1)/2$, depending on the product of the diffraction order $m$ and the grating singularity order $p$. Whereas, in [3] the singularity order of the diffracted beam is equal to $n$, and the beam is described by Bessel functions of order $(n-1)/2$ and $(n+1)/2$, depending on the phase layer singularity $n$.

The vortex wave field, described by a difference of two modified Bessel functions of fractional order $(m \pm 1)/2$, according to D. Rozas et al., in [22] is a result of interaction between the vortex beam and the background field. Free space paraxial propagation of an array of vortices, nested in a Gaussian beam, and each of form $r^m \exp(im\theta)\exp(-r^2/w_0^2)$, has been investigated in [23] showing that vortices of opposite charges attract each other and can collide and annihilate, whereas, if they are of same topological charge, the array simply expands or contracts in the host beam and rotates rigidly. Their Fourier transform is expressed as a product of a Kummer function and a donut-Gaussian function profile.

In order to find the dimensions of the vortices of the beams (19) we use the radial intensity distribution. To find it, it is necessary to make a separation of the real and the imaginary part in the expressions (19), which is not possible because of the complex argument $ik\rho_{\pm1}^2/2Q(z)$ of the Bessel functions. For that purpose, we use the Neumann's addition theorem (see appendix 2) and we can define the higher diffraction order beams through the Bessel functions of real arguments, at any position from the grating plane. The derivation of the intensity upon the radial coordinate and making it equal to zero (in order to find the vortex radii), results in a transcendental equation which admits no analytical solutions for arbitrary charge order $p$, but can be solved only numerically for particular values of $m$ and $p$, if needed.

Some general information about the spatial behavior of the vortex beams produced by the forked gratings and vortex radii can be obtained for very near and far field.

## 4. THE NEAR-FIELD APPROXIMATION

When $(z-\zeta)\to 0$ and $\rho_{\pm1}\to 0$, the approximate value of the Bessel functions argument is

$$\frac{k\rho_{\pm m}^2}{2Q(z)} \to -\frac{k\rho_{\pm m}^2}{4(z-\zeta)}.$$

We took into account that $R(z)\approx R(\zeta) \underset{\rho\to 0}{\to} \infty$; $w(z)\approx w(\zeta) \gg \rho_{\pm m}$ and $\exp\left[-\rho_{\pm m}^2/w^2(\zeta)\right]\to 1$.

From equations (22) the wave amplitudes are obtained in the form

$$U_{\pm m}(\rho_{\pm m},\theta_{\pm m},z)=\frac{t_{\pm m}}{2}\frac{q(0)}{q(\zeta)}\left[\frac{-ik\pi}{2(z-\zeta)}\right]^{1/2}(\pm 1)^{mp}(-i)^{(mp-1)/2}\exp\left\{-ik\left[z+\frac{\rho^2}{2(z-\zeta)}-\frac{\rho_{\pm m}^2}{4(z-\zeta)}\right]\right\}$$

$$\times \exp\left[imp\left(\frac{\pi}{2}\pm\theta_{\pm m}\right)\right]\rho_{\pm m}\left[J_{\frac{mp-1}{2}}\left(\frac{k\rho_{\pm m}^2}{4(z-\zeta)}\right)+iJ_{\frac{mp+1}{2}}\left(\frac{k\rho_{\pm m}^2}{4(z-\zeta)}\right)\right], \qquad (25)$$

and their intensity distributions as

$$\hat{I}_{\pm m}(\rho_{\pm m},\theta_{\pm m},z)=|t_{\pm m}|^2\frac{w_0^2}{w^2(\zeta)}\frac{k\pi}{8(z-\zeta)}\rho_{\pm m}^2\left[J_{\frac{mp-1}{2}}^2\left(\frac{k\rho_{\pm m}^2}{4(z-\zeta)}\right)+J_{\frac{mp+1}{2}}^2\left(\frac{k\rho_{\pm m}^2}{4(z-\zeta)}\right)\right]. \qquad (26)$$

Further, we find the intensity first derivatives upon the radial coordinates $\rho_{\pm m}$, make them equal to zero, and using some relations for Bessel functions we find that, besides at vortex zeros $\rho_{\pm m} = 0$, the corresponding derivatives are equal to zero when

$$\left| J_{\frac{mp-1}{2}}\left(\frac{k\rho_{\pm m}^2}{4(z-\zeta)}\right) \right| = \left| J_{\frac{mp+1}{2}}\left(\frac{k\rho_{\pm m}^2}{4(z-\zeta)}\right) \right|. \qquad (27)$$

From the tables of Bessel functions we can approximate the first zero of Eq. (27) as $k\rho_{\pm m}^2 / 4(z-\zeta) \approx mp/2 + 1$, $(mp = 1, 2, ... < 30)$, or the vortex radii in the near field are

$$(\rho_{\pm m})_p \approx \left[\left(\frac{mp}{2}+1\right)\frac{2\lambda(z-\zeta)}{\pi}\right]^{1/2}. \qquad (28)$$

At a given small distance value of (z-ζ), the higher charged order vortices are wider. For a singularity order *mp*, the maximums of the light rings around the vortices are found on the rotational paraboloids. Considering the small argument approximation for the Bessel functions

$$J_\nu(x) = \left(\frac{x}{2}\right)^\nu \frac{1}{\Gamma(\nu+1)} \text{ when } x \to 0,$$

it is found that the intensity drops down to zero near the vortex axes as

$$\hat{I}_{\pm m}(\rho_{\pm m}, \theta_{\pm m}, z) \approx \Gamma^{-2}\left(\frac{mp+1}{2}\right)\left[\frac{k\rho_{\pm m}^2}{8(z-\zeta)}\right]^{(mp-1)}\left[1 + \frac{k^2\rho_{\pm m}^4}{16(mp+1)^2(z-\zeta)^2}\right]. \qquad (29)$$

It means that in a given diffraction order, at a given distance (z-ζ), the increasing of the vortex charge order makes the walls of the vortex to become more vertical. This conclusion can be also seen in the Fig. 4, where the plotted graphs are based on the analytical solution (26) for the near-field intensity. In the Fig. 5 the radial intensity distribution around the first diffraction order vortex is shown for different small distances $z-\zeta$. In the both graphs, the oscillating behavior of the radial intensity around the value one, when going away from the vortex center, is clearly seen (at bigger radial distances from the vortex center, this ringing intensity is decreasing to zero). The parameters used are: $w_0 = 1$ mm, $\lambda = 800$ nm. The plotted graphs are based on Eqs. (26).

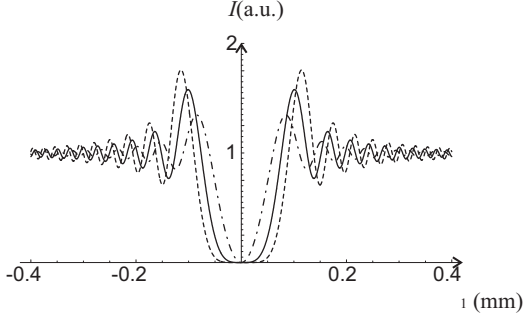
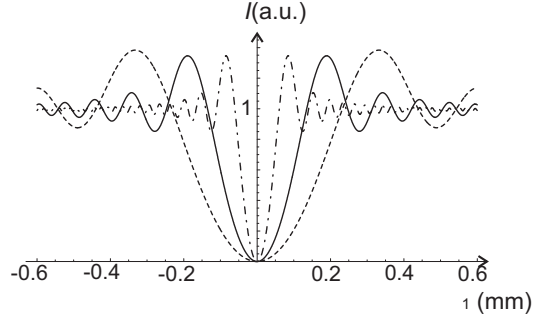

Fig. 4. Intensity distribution in the first diffraction order for diffraction of a Gaussian beam by a forked grating with singularity: $p=1$ (dot-dashed line), $p=2$ (full line), $p=3$ (dashed line) at distance $z-\zeta=10$ mm.

Fig. 5. Intensity distribution in the first diffraction order for diffraction of a Gaussian beam by a forked grating with singularity $p=1$ at distances: $z-\zeta=10$ mm (dot-dashed line), $z-\zeta=50$ mm (full line), $z-\zeta=150$ mm (dashed line).

## 5. THE FAR-FIELD APPROXIMATION

When $(z-\zeta)\to\infty$, then we can approximate

$$\frac{1}{Q(z)}=\frac{1}{2}\left\{\left[\frac{1}{R(z)}-\frac{1}{z-\zeta}\right]-\frac{2i}{kw'^2(z)}\right\}\to\frac{-i}{kw^2(z)}.$$

According to the expressions (19), the waves with phase singularities can be approximated in the form

$$U_{\pm m}(\rho_{\pm m},\theta_{\pm m},z)=\frac{t_{\pm m}}{2}\frac{q(0)}{q(z)}\frac{\sqrt{\pi}}{w(z)}(\pm 1)^{mp}\exp(-ikz)\exp\left[-\frac{\rho_{\pm m}^2}{2w^2(z)}\right]\exp\left[imp\left(\frac{\pi}{2}\pm\theta_{\pm m}\right)\right]$$

$$\times \rho_{\pm m}\left[I_{\frac{mp-1}{2}}\left(\frac{\rho_{\pm m}^2}{2w^2(z)}\right)-I_{\frac{mp+1}{2}}\left(\frac{\rho_{\pm m}^2}{2w^2(z)}\right)\right], \tag{30}$$

and, accordingly, their intensities are

$$\hat{I}_{\pm m}(\rho_{\pm m},\theta_{\pm m},z)=\frac{w_0^2}{w^2(z)}|t_{\pm m}|^2\frac{\pi}{4}\frac{\rho_{\pm m}^2}{w^2(z)}\exp\left[-\frac{\rho_{\pm m}^2}{w^2(z)}\right]\left[I_{\frac{mp-1}{2}}\left(\frac{\rho_{\pm m}^2}{2w^2(z)}\right)-I_{\frac{mp+1}{2}}\left(\frac{\rho_{\pm m}^2}{2w^2(z)}\right)\right]^2. \tag{31}$$

Expressions (31) are similar to the far field intensity distribution of a Gaussian beam diffracted by spiral phase plate of topological charge $n$ in [3], but they differ in the orders of the Bessel functions. Here, they are equal to $(mp\mp 1)/2$, depending on the product of the diffraction order $m$ and the grating singularity order $p$, instead of $(n\mp 1)/2$ in [3], depending on the phase layer singularity $n$. Also, the argument of the Bessel functions here is $\rho_{\pm m}^2/2w^2(z)$, instead of $\rho_{\pm m}^2/w^2(z)$.

In the beam paraxial region we utilize the small argument approximation for the modified Bessel functions, approximate $\exp[-\rho_{\pm m}^2/w^2(z)] \approx 1$, and find that

$$\hat{I}_{\pm m}(\rho_{\pm m}, \theta_{\pm m}, z) \approx \Gamma^{-2}\left(\frac{mp+1}{2}\right)\left[\frac{1}{4w^2(z)}\right]^{mp} \rho_{\pm m}^{2mp} \exp\left[-\frac{\rho_{\pm m}^2}{w^2(z)}\right]\left[1 - \frac{\rho_{\pm m}^2}{2(mp+1)w^2(z)}\right]^2. \quad (32)$$

The vortex radii are found as

$$(\rho_{\pm m})_p = w'(z)\sqrt{\frac{mp(mp+1)}{mp+2}}. \quad (33)$$

Far from the singularity axis, when $\rho_{\pm m} \to \infty$, we use the large argument approximation for the modified Bessel functions

$$I_\nu(x) \approx \frac{1}{\sqrt{2\pi x}}\exp(x)\left(1 - \frac{4\nu^2 - 1}{8x}\right)$$

and find that the intensity decreases as

$$\hat{I}_{\pm m}(\rho_{\pm m} \to \infty, z) \propto |t_{\pm m}|^2 \frac{w_0^2 w^2(z)}{8} \frac{(mp)^2}{\rho_{\pm m}^4}. \quad (34)$$

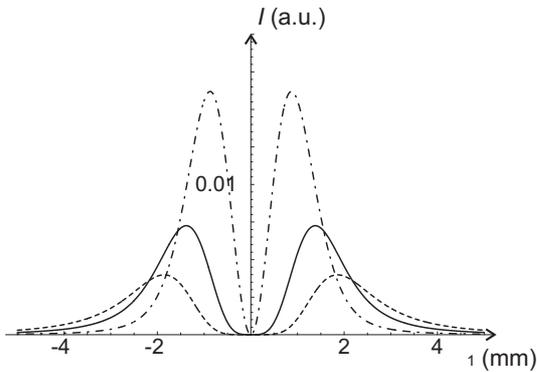
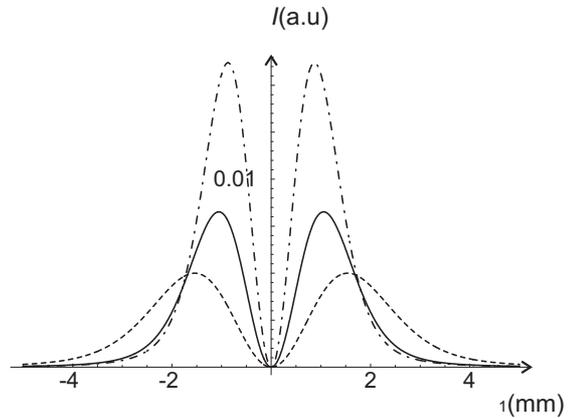

Fig. 6. Intensity distribution in the first diffraction order for diffraction of a Gaussian beam by forked grating with singularity: *p*=1 (dot-dashed line), *p*=2 (full line), *p*=3 (dashed line), at distance *z*=1 m.

Fig. 7. Intensity distribution in the first diffraction order for diffraction of a Gaussian beam by forked grating with singularity *p*=1 at distances: *z*=1 m (dot-dashed line), *z*=3 m (full line), *z*=5 m (dashed line).

In the Fig. 6 the radial intensity distribution around the vortex in the first diffraction order, at distance *z*=1 m, and for different topological charges of the grating is shown. Whereas, in the Fig. 7, the analog plots are made in order to show the intensity decreasing along the z axis, for a given value of the topological charge. The plotted graphs are based on Eqs. (31).

## 6. SPECIALIZATION OF THE RESULTS FOR THE CASE $p=0$

It was previously agreed that the general solution concerns $p=0$ as a special case i.e. when the grating is rectilinear instead of being forked.

We use Eq. (19) and specialize it for the value $p=0$, applying also the identity

$$I_{\pm\frac{1}{2}}(x) = \sqrt{\frac{2}{\pi x}} \begin{cases} \operatorname{sh}x \\ \operatorname{ch}x \end{cases} = \sqrt{\frac{2}{\pi x}} \begin{cases} [\exp(x) - \exp(-x)]/2 \\ [\exp(x) + \exp(-x)]/2 \end{cases},$$

and get the higher diffraction order wave amplitudes in the form

$$U_{\pm m}(\rho_{\pm m}, \theta_{\pm m}, z) = t_{\pm m} \frac{q(0)}{q(z)} \exp\left\{-ik\left[z + \frac{\rho^2 - \rho_{\pm m}^2}{2(z-\zeta)}\right]\right\} \exp\left[-\frac{ik}{2q(z)}\rho_{\pm m}^2\right], \tag{35}$$

and the corresponding intensities

$$\hat{I}_{\pm m}(\rho_{\pm m}, \theta_{\pm m}, z) = |t_{\pm m}|^2 \frac{w_0^2}{w^2(z)} \exp\left[-\frac{2\rho_{\pm m}^2}{w^2(z)}\right]. \tag{36}$$

When $p=0$, the zero diffraction order wave amplitude is

$$U_0(\rho, z) = t_0 \frac{q(0)}{q(z)} \exp\left\{-ik\left[z + \frac{\rho^2}{2q(z)}\right]\right\}, \tag{37}$$

and the corresponding intensity is

$$\hat{I}_0(\rho, z) = |t_0|^2 \frac{w_0^2}{w^2(z)} \exp\left(-\frac{2\rho^2}{w^2(z)}\right). \tag{38}$$

From the Eqs. (36) and (38) it can be concluded easily that the axial intensities $\hat{I}_0(\rho=0, \theta, z), \hat{I}_{\pm m}(\rho_{\pm m}=0, \theta_{\pm m}, z)$ are different from zero, as it is expected.

## 7. FRAUNHOFER DIFFRACTION OF A GAUSSIAN BEAM ON A FSG

We consider a scalar diffraction of a Gaussian beam

$$U^{(i)}(r, \varphi) = A\exp\left(-\frac{r^2}{w_o^2}\right), \tag{39}$$

where $w_0$ is the radius of the Gaussian beam waist, A is a constant amplitude value in the beam center, by a forked grating whose transmittance is given by Eq. (1), incident with its waist in the grating plane.

In the focal plane of a convergent lens with focal distance $f$, the intensity distribution can be found using the diffraction integral [19]

$$U(\rho,\theta,f) = C \iint_\sigma T(r,\varphi) U^{(i)}(r,\varphi) \exp\left(\frac{ik\rho}{f} r \cos(\varphi - \theta)\right) r \, dr \, d\varphi \;, \tag{40}$$

where $C = i/\lambda f$ is a complex constant. The integration in Eq. (40) is over the area of the grating, but, since for $r > w_0$, $U^{(i)} \to 0$, the integration over the radial variable can be performed in the interval $[0, \infty]$. In the process of integration we are using similar transformations for the variables in the focal plane as those given by Eq. (6)

$$\rho \cos\theta \mp m\frac{\lambda f}{D} = \rho_{\pm m} \cos\theta_{\pm m}; \quad \rho \sin\theta = \rho_{\pm m} \sin\theta_{\pm m};$$

$$\rho_{\pm m} = \sqrt{\rho^2 + \left[\frac{m\lambda f}{D}\right]^2 \mp \frac{2m\lambda \rho f}{D} \cos\theta}; \quad \text{tg}\,\theta_{\pm m} = \frac{\rho \sin\theta}{\rho \cos\theta \mp m\lambda f/D}.$$

The final expressions for the diffracted wave fields in the Fourier plane are

$$U_0(\rho, f) = A\frac{iw_0}{w_f} t_0 \exp\left(-\frac{\rho^2}{w_f^2}\right) \tag{41}$$

for the zero-th diffraction order, and

$$U_{\pm m}(\rho_{\pm m}, \theta_{\pm m}, f) = A(\pm 1)^{mp} t_{\pm m} \sqrt{\frac{\pi}{2}} \frac{w_0}{w_f} \exp[imp(\pi/2 \pm \theta_{\pm m})] \frac{\rho_{\pm m}}{\sqrt{2} w_f}$$

$$\times \exp\left[-\frac{\rho_{\pm m}^2}{2w_f^2}\right]\left[I_{\frac{mp-1}{2}}\left(\frac{\rho_{\pm m}^2}{2w_f^2}\right) - I_{\frac{mp+1}{2}}\left(\frac{\rho_{\pm m}^2}{2w_f^2}\right)\right] \tag{42}$$

for the higher diffraction orders, where we have denoted

$$w_f = \frac{\lambda f}{w_0 \pi}. \tag{43}$$

From the Eq. (41) it is seen that the zero-th diffraction order field is represented by a Gaussian bright spot with radius $w_f = \lambda f / w_o \pi$ and intensity

$$\hat{I}_0(\rho, f) = A^2 |t_0|^2 \frac{w_0^2}{w_f^2} \exp\left(-\frac{2\rho^2}{w_f^2}\right). \tag{44}$$

The wave fields (42) are similar to those obtained for the far-field approximation of Fresnel diffraction of a Gaussian beam by forked grating, with these differences:

-The centers of the components $U_{\pm m}(\rho_{\pm m}, \theta_{\pm m}, f)$ in the focal plane, where the vortices exist, are the points $C_m(m\lambda f/D, 0)$ and $C_{-m}(m\lambda f/D, \pi)$. Two near-by vortex beams are separated by distance $C_{m-1}C_m = \lambda f/D$.

- In the argument of the Bessel functions, now, instead of $w(z) = w_0\sqrt{1 + (2z/kw_0^2)^2}$ we have $w_f$, given by Eq. (43).

-From the intensity radial distribution in the *m*-th diffraction order, obtained from the Eq. (42),

$$\hat{I}_{\pm m}(\rho_{\pm m}, \theta_{\pm m}, f) = A^2 \frac{\pi}{2}|t_{\pm m}|^2 \left(\frac{w_0}{w_f}\right)^2 \left(\frac{\rho_{\pm m}}{\sqrt{2}w_f}\right)^2 \exp\left[-\frac{\rho_{\pm m}^2}{w_f^2}\right] \left[I_{\frac{mp-1}{2}}\left(\frac{\rho_{\pm m}^2}{2w_f^2}\right) - I_{\frac{mp+1}{2}}\left(\frac{\rho_{\pm m}^2}{2w_f^2}\right)\right]^2, \tag{45}$$

the vortex radii are found in a similar way as it has been done previously in section 5, and are given by the expression

$$\rho_{\pm m} = w_f \sqrt{2}\sqrt{\frac{mp(mp+1)}{mp+2}} = \frac{\lambda f}{w_0 \pi}\sqrt{2}\sqrt{\frac{mp(mp+1)}{mp+2}}. \tag{46}$$

In [6] the solution of a Fraunhofer diffraction of a Gaussian beam by a helical axicon is deduced in a form of a series of confluent hypergeometric functions, and in a form of finite sum of the Bessel functions of orders $(n \pm 1)/2$ for a phase spiral plate of singularity order *n* (helical axicon transforms into phase spiral plate when the axicon angle tends to zero). In both cases the diffracted beam carries topological charge equal to the singularity order *n* of the spiral phase plate.

## 8. CONCLUSIONS

From the theoretical results of this article we can draw out the following conclusions:

-The fork-shaped grating with *p* internal "teeth" diffracts the Gaussian normally incident beam into a fan of beams, whose number depends on the type of the grating. The amplitude reduced straight-through beam is a Gaussian (for ideal binary phase grating it is absent), while the higher diffraction order beams, deviated with respect to the incident direction, are hollow beams.

-The axes of the tilted beams are dark, due to the phase singularities of order $+mp$ and $-mp$, indicating the opposite chiralities of their wavefronts-they are vortex beams.

-The radial parts of the wave functions, describing the spatial form of the vortex beams are not of the "nondiffracting" integer order Bessel type, nor of the Laguerre-Gaussian type. They are described by a doughnut Gaussian shape, additionally "complicated" by terms consisted of Bessel functions and modified Bessel functions of real argument, whose orders don't match the value of the phase singularity charge of the vortices. They are of the spiral phase plate type.

-The dark spaces around the vortex axes are wrapped round by light rotational hyperboloids. Their presence in the observation plane is registered by light rings of radii given by equations (28) and (33) for near and far field, respectively. The rings are around dark spots with centres $C_m(m\lambda(z-\zeta)/D, 0)$ and $C_{-m}(m\lambda(z-\zeta)/D, \pi)$.

-From its maximum value on the ring, the intensity drops down to zero towards the vortex axes (vortex centers $C_{\pm m}$ in the observation plane) as $\rho_{\pm m}^{2mp}$ in both near and far field approximation.

-On the other side from the light rings, in the near field approximation the intensity drop tends to one, while in the far field it drops down to zero like $(mp)^2/\rho_{\pm m}^4$.

-The equations $\Psi(\rho_{\pm m}, \theta_{\pm m}) - \Psi_{\pm m}(\rho_{\pm m}, \theta_{\pm m}) = n\pi$ ($|n| = 0,1,2,..$), where $\Psi(\rho_{\pm m}, \theta_{\pm m}) = (2\pi/D)\rho_{\pm m}\cos\theta_{\pm m}$ is a phase function of a plane wave defined in polar coordinate systems with $C_{\pm m}$ taken as poles respectively, and $\Psi_{\pm m}(\rho_{\pm m}, \theta_{\pm m})$ are the phase functions of the vortex beams, define the fringes of their interference pattern. Realized in the plane $z$=constant, in the far field of the vortex beams $U_{\pm m}$, the fringe pattern is defined by

$$\frac{2\pi}{D'}\rho_{\pm m}\cos\theta_{\pm m} \mp mp\theta_{\pm m} = n\pi; \quad |n| = 0,1,2,..$$

It is the fork-shaped fringe pattern, with $mp$ internal 'teeth", as it was shown in the experiments reported by Heckenberg et al. in [8].

APPENDIX 1

Let us use the following notation for the complex argument of Kummer function in Eq. (17)

$$2y_{\pm m} = -\frac{ik}{2}\frac{q(\zeta)}{(z-\zeta)q(z)}\rho_{\pm m}^2. \qquad (1\text{ A})$$

Among the variety of recurrence relations satisfied by the Kummer functions, we use the relation [21]

$$(\nu-1)M(\mu,\nu-1;2y_{\pm m}) = (\nu-1)M(\mu,\nu;2y_{\pm m}) + 2y_{\pm m}\frac{\mu}{\nu}M(\mu+1,\nu+1;2y_{\pm m}), \qquad (2\text{ A})$$

and adapting it to the values $\mu = mp/2, \nu = mp+1$, get

$$M\left(\frac{mp}{2},mp+1;2y_{\pm m}\right) = M\left(\frac{mp}{2},mp;2y_{\pm m}\right) - \frac{y_{\pm m}}{mp+1}M\left(\frac{mp}{2}+1,mp+2;2y_{\pm m}\right). \qquad (3\text{ A})$$

The right side of this expression is related to the difference of two modified Bessel functions. Taking into account that

$$I_{\frac{mp-1}{2}}(y_{\pm m}) = \Gamma^{-1}\left(\frac{mp+1}{2}\right)\left(\frac{y_{\pm m}}{2}\right)^{(mp-1)/2}\exp(-y_{\pm m})M\left(\frac{mp}{2},mp;2y_{\pm m}\right) \qquad (4\text{ A})$$

and

$$I_{\frac{mp+1}{2}}(y_{\pm m}) = \Gamma^{-1}\left(\frac{mp+1}{2}\right)\left(\frac{y_{\pm m}}{2}\right)^{(mp-1)/2}\exp(-y_{\pm m})\frac{y_{\pm m}}{(mp+1)}M\left(\frac{mp}{2}+1,mp+2;2y_{\pm m}\right), \qquad (5\text{ A})$$

we find that

$$M\left(\frac{mp}{2},mp+1;2y_{\pm m}\right) = \Gamma\left(\frac{mp+1}{2}\right)\left(\frac{y_{\pm m}}{2}\right)^{(1-mp)/2}\exp(y_{\pm m})\left[I_{\frac{mp-1}{2}}(y_{\pm m}) - I_{\frac{mp+1}{2}}(y_{\pm m})\right], \qquad (6\text{ A})$$

APPENDIX 2

Since the beams (16) and (19) are Gaussian modulated, only at small distances $(z-\zeta)$ behind the grating they overlap each other and interfere. At bigger distances they exist separately, and the intensity distribution can be calculated separately for each diffracted component. The intensities of the beams are

$$\hat{I}_0(\rho,z) = |U_0(\rho,z)|^2 = |t_0|^2\frac{w_0^2}{w^2(z)}\exp\left[\frac{-2\rho^2}{w^2(z)}\right], \qquad (7\text{ A})$$

$$\hat{I}_{\pm m}(\rho_{\pm m},\theta_{\pm m},z)=\left|U_{\pm m}(\rho_{\pm m},\theta_{\pm m},z)\right|^2=\left|t_{\pm m}\right|^2\frac{w_0^2}{w^2(z)}K(z)\rho_{\pm m}^2\exp\left[\frac{-\rho_{\pm m}^2}{w^2(z)}\right]\left[A_{mp}^2(\rho_{\pm m},z)+B_{mp}^2(\rho_{\pm m},z)\right]$$
(8 A)

In (8 A) we have denoted as

$$K(z)=\frac{\pi}{2}\left[\left(\frac{k}{2R'(z)}\right)^2+\left(\frac{1}{w'^2(z)}\right)^2\right],$$
(9 A)

while $A_{mp}(\rho_{\pm m},z)$ and $B_{mp}(\rho_{\pm m},z)$ are functions of real arguments, obtained by application of Neumann's addition theorem [21]

$$J_\nu(u-g)=\sum_{s=-\infty}^{\infty}J_{\nu+s}(u)J_s(g) \quad \text{when} \quad \left|\frac{g}{u}\right|<1,$$

$$J_\nu(u-g)=(-1)^\nu J_\nu(g-u)=(-1)^\nu\sum_{s=-\infty}^{\infty}J_{\nu+s}(g)J_s(u), \quad \text{when} \quad \left|\frac{u}{g}\right|<1,$$
(10 A)

to the Bessel functions $J_{\frac{mp\mp 1}{2}}\left(\left(\frac{k}{2R'(z)}-\frac{1}{w'^2(z)}\right)\rho_{\pm m}^2\right)$ in Eq. (22).

When

$$\frac{k}{2R'(z)}<\frac{1}{w'^2(z)},$$

or, for distances
$$z-\zeta<\frac{kw^2(z)}{kw^2(z)/R(z)-2},$$
(11 A)

the coefficients $A_{mp}(\rho_{\pm m},z)$ and $B_{mp}(\rho_{\pm m},z)$ are

$$A_{mp}(\rho_{\pm m},z)=\sum_{s'=-\infty}^{\infty}(-1)^{s'}J_{2s'}\left(\frac{k\rho_{\pm m}^2}{2R'(z)}\right)\left[I_{\frac{mp-1}{2}+2s'}\left(\frac{\rho_{\pm m}^2}{w'^2(z)}\right)-I_{\frac{mp+1}{2}+2s'}\left(\frac{\rho_{\pm m}^2}{w'^2(z)}\right)\right],$$
(12 A)

$$B_{mp}(\rho_{\pm m},z)=\sum_{s'=-\infty}^{\infty}(-1)^{s'}J_{2s'+1}\left(\frac{k\rho_{\pm m}^2}{2R'(z)}\right)\left[I_{\frac{mp-1}{2}+2s'+1}\left(\frac{\rho_{\pm m}^2}{w'^2(z)}\right)-I_{\frac{mp+1}{2}+2s'+1}\left(\frac{\rho_{\pm m}^2}{w'^2(z)}\right)\right].$$
(13 A)

Whereas, when

$$\frac{k}{2R'(z)}>\frac{1}{w'^2(z)},$$

or, for distances
$$z-\zeta>\frac{kw^2(z)}{kw^2(z)/R(z)-2},$$
(14 A)

they are given as

$$A_{mp}(\rho_{\pm m},z)= \sum_{s'=-\infty}^{\infty} (-1)^{s'}\left[ I_{2s'}\left(\frac{\rho_{\pm m}^2}{w'^2(z)}\right) J_{\frac{mp-1}{2}+2s'}\left(\frac{k\rho_{\pm m}^2}{2R'(z)}\right) + I_{2s'+1}\left(\frac{\rho_{\pm m}^2}{w'^2(z)}\right) J_{\frac{mp+1}{2}+(2s'+1)}\left(\frac{k\rho_{\pm m}^2}{2R'(z)}\right)\right], \quad (15\ A)$$

$$B_{mp}(\rho_{\pm m},z)= \sum_{s'=-\infty}^{\infty} (-1)^{s'}\left[ I_{2s'+1}\left(\frac{\rho_{\pm m}^2}{w'^2(z)}\right) J_{\frac{mp-1}{2}+(2s'+1)}\left(\frac{k\rho_{\pm m}^2}{2R'(z)}\right) - I_{2s'}\left(\frac{\rho_{\pm m}^2}{w'^2(z)}\right) J_{\frac{mp+1}{2}+2s'}\left(\frac{k\rho_{\pm m}^2}{2R'(z)}\right)\right]. \quad (16\ A)$$

The doughnut Gaussian shapes of the beams (8 A) are additionally "complicated" by the terms $\left[A_{mp}^2(\rho_{\pm m},z)+ B_{mp}^2(\rho_{\pm m},z)\right]$, which take part in the determination of the vortex core sizes of the beams. Being functions of the distance $z$, they, together with the variable Gaussian cross section radius $w(z)$, clearly indicate that vortex size in a given diffraction order depends on the distance $z$ and on the charge order $p$.

Bringing to zero the first derivative of expressions (8 A) with respect to the variables $\rho_{\pm m}$ will give the equations for finding the intensity extremes

$$\frac{d\hat{I}_{\pm m}(\rho_{\pm m},\theta_{\pm m},z)}{d\rho_{\pm m}} = 2K(z)\rho_{\pm m} \exp\left[\frac{-\rho_{\pm m}^2}{w^2(z)}\right]\left\{\left(1-\frac{\rho_{\pm m}^2}{w^2(z)}\right)\left[A_{mp}^2(\rho_{\pm m},z)+ B_{mp}^2(\rho_{\pm m},z)\right]\right.$$
$$\left. + \rho_{\pm m}\left[A_{mp}(\rho_{\pm m},z)\frac{dA_{mp}(\rho_{\pm m},z)}{d\rho_{\pm m}}+ B_{mp}(\rho_{\pm m},z)\frac{dB_{mp}(\rho_{\pm m},z)}{d\rho_{\pm m}}\right]\right\} = 0. \quad (17\ A)$$

The solutions $\rho_{\pm m}=0$ give the vortex zero. For our purpose, of importance are the first maxima which define the vortex size. The key for their identification lays in the solution of the equation

$$\left(1-\frac{\rho_{\pm m}^2}{w^2}\right)\left[A_{mp}^2(\rho_{\pm m},z)+ B_{mp}^2(\rho_{\pm m},z)\right] + \rho_{\pm m}\frac{d}{d\rho_{\pm m}}\left[A_{mp}^2(\rho_{\pm m},z)+ B_{mp}^2(\rho_{\pm m},z)\right] = 0. \quad (18\ A)$$

With $A_{mp}(\rho_{\pm m},z)$ and $B_{mp}(\rho_{\pm m},z)$ as are defined in Eqs. (12 A), (13 A), (15 A), (16 A), equations (18 A) are rather transcendental and admit no analytical solutions for arbitrary charge order $p$. They can be solved only numerically for particular values of $m$ and $p$.